\documentclass[pra,showpacs,amsmath,amssymb,twocolumn]{revtex4}
\usepackage{amssymb}
\usepackage{mathrsfs}
\usepackage{amsfonts}

\usepackage{amsthm}
\usepackage{dcolumn}
\usepackage{bm}
\usepackage{graphicx}
\pacs{03.67.Pp, 03.65.Ud}
\newcommand\ket[1]{\ensuremath{|#1\rangle}}
\newcommand\bra[1]{\ensuremath{\langle#1|}}

\begin{document}
\title{Protecting Two-Qubit Quantum States by $\pi$-Phase Pulses}
\author{Jia-Zhong Hu}
  \affiliation{Department of Physics and  the Key Laboratory of Atomic
and Nanosciences, Ministry of Education, Tsinghua University,
Beijing 100084, China}
\author{Xiang-Bin Wang}
  \email{xbwang@mail.tsinghua.edu.cn}
  \affiliation{Department of Physics and  the Key Laboratory of Atomic
and Nanosciences, Ministry of Education, Tsinghua University,
Beijing 100084, China}
\author{Leong Chuan Kwek}
\email{kwekleongchuan@nus.edu.sg}
  \affiliation{Certer for Quantum Technologies, National University of Singapore, 2 Science Driver 3, Singapore 117542}
  \affiliation{National Institute of Education and Institute of Advanced Studies, Nanyang Technological University,
  1 Nanyang Walk, Singapore 637616}

\begin{abstract}
We study the state decay of two qubits interacted with a common
harmonic oscillator reservoir. There are both decoherence error and
the error caused by the amplitude change of the superradiant state.
We show that frequent $\pi$-phase pulses can eliminate both typpes
of errors therefore protect a two-qubit odd-parity state more
effectively than the frequent measurement method. This shows that the the methods using dynamical decoupling and 
the quantum Zeno effects actually can give rather {\em different} results when the operation frequency is finite.
\end{abstract}

\maketitle

\section{introduction}
The interaction between a quantum system and its environment
inevitably leads to the decoherence\cite{Peter,Kofman} of a quantum
state.  Such quantum decoherence can often cause severe distortion
to a quantum state rendering many quantum systems in the real world
useless\cite{EPR,QTEL,DiVincenzo,DiVi,Josephson,Huang,Duan,Hu}. In
order to protect a quantum state, many methods against decoherence
have been studied. Among the existing proposals, most of them are
for single-qubit state
protection\cite{Viola,Breuer,Vitali,Fanchini,Uhrig,Yang}.
Recently\cite{Man},  a scheme for the protection of quantum
entanglement of  two qubits at 0K temperature was proposed using
quantum Zeno effect (QZE), i.e.,  via frequent measurement of the
environment photon number for the Jaynes-Cummings (J-C) model.
However, as shown below,  besides the decoherence error, the
amplitude of the superradiant state can also changed. It decreases
with evolution. Intuitively speaking, the superradiant state changes
into $|00\rangle$ gradually in  the evolution therefore the initial
odd-parity state cab be severely distorted after a long time
evolution. The frequent measurement method cannot eliminate such
distortion efficiently because it actually removes the term
$|00\rangle$ at every step. After a long evolution time with a fixed
measurement frequency, the amplitude of the superradiant state
decreases a lot therefore severely distort the initial unknown
state. On the other hand, given the existing technologies, it seems
that the measurement of photon number of the environment of all
modes remains a challenging task. It is therefore an interesting
problem to study how one could protect a two-qubit state with
techniques which have been demonstrated already, for example,
dynamical decoupling
scheme\cite{Viola,Breuer,Vitali,Fanchini,Uhrig,Yang,West} which have
been demonstrated experimentally recently\cite{Du,Biercuk,Uys}. It is well known that in
the limit of infinitely frequent operations, QZE and dynamical decoupling
are unified and can have the same
results\cite{Facchi1,Facchi2,Busch}. The two methods are not compared in the more realistic condition when
the operation frequencies are finite.
 Here we show that
the dynamical decoupling scheme achieved through a frequent
application of $\pi$-phase pulses can protect a two-qubit state more
effectively. The scheme  not only protects the state from
decoherence error, but also prevents the amplitude changing of
superradiant state.

This paper is arranged as following:  we
first review the existing results of the   J-C
model\cite{Scully,Man} for two qubits,  in particular,  the time
evolution of the odd parity state\cite{Man} under zero temperature.
We point out why the amplitude of superradiant state changes in a
frequent measurement scheme. We then show how to protect the state
by $\pi-$phase pulses and why $\pi-$phase pulses scheme can prevent
the amplitude change. The consequences of the finite frequency and
duration of each pulses are also presented.

\section{Amplitude change of supperradiant  state in J-C model}
Consider the following Hamiltonian for a two-qubit system and its
environment as used in\cite{Man}:
\begin{equation}
H=H_s+H_e+H_i
\end{equation}
where
\begin{equation}
H_s=\omega_0
(\sigma^{+}_{1}\sigma^{-}_{1}+\sigma^{+}_{2}\sigma^{-}_{2})
\end{equation}
is Hamiltonian of the two-qubits (system),
\begin{equation}
H_e=\sum_k\omega_{k} b^+_{k}b_{k}
\end{equation}
is the Hamiltonian of the environment, and
\begin{equation}
H_i=(\alpha_{1}\sigma^{+}_{1}+\alpha_{2}\sigma^{+}_{2})\sum_k
g_{k}b_{k}+h.c.
\end{equation}
 is Hamiltonian of the interaction between the system and the environment. Notations $b_k$ and
$b^+_k$ are the annihilation and creation operators of the
environment with frequency $\omega_{k}$; $\omega_0$ is the atomic
transition frequency between the ground state $\ket{0}$ and the
excited state $\ket{1}$; $\sigma^{+}=\ket{1}\bra{0}$ and
$\sigma^{-}=\ket{0}\bra{1}$. The solution of such a model for the case of zero environmental
temperature is well known and it can be
found in Ref. \cite{Scully,Tavis}. In particular, for an odd-parity two-qubit initial
state, there exists a dark state \begin{equation}\ket{\mu}={\alpha_2
\over \alpha}\ket{1}_1\ket{0}_2-{\alpha_1 \over
\alpha}\ket{0}_1\ket{1}_2
\end{equation}and a superradiant
state\cite{Yu,Almeida,Palma,Zanardi}.
\begin{eqnarray}
 \ket{\nu}={\alpha_1 \over
\alpha}\ket{1}_1\ket{0}_2+{\alpha_2 \over \alpha}\ket{0}_1\ket{1}_2.
\end{eqnarray}
The dark state $|\mu\rangle$ does not change with time under the JC
model, while the  superradiant state changes with time according to
\begin{equation}
\ket{\nu}\otimes\ket{0}_e
=\eta(t)\ket{\nu}\otimes\ket{0}_e+\ket{0}_1\ket{0}_2\otimes
\sum_k\left(c(t)_k\ket{1_k}_e\right) \label{timesup}
\end{equation} and
\begin{equation}
\eta(t)=e^{-{\lambda t\over 2}}[\cosh({\Omega t \over 2})+{\lambda
\over \Omega}\sinh ({\Omega t \over 2})]\label{eta}
\end{equation}
where $\Omega=\sqrt {\lambda^2-4R^2}$, $\alpha=\sqrt
{\alpha^2_1+\alpha^2_2}$ and  $R=\alpha W$. Given any odd-parity
initial state $|\psi_0\rangle = \beta_1|\mu\rangle +
\beta_2|\nu\rangle$, the time evolution is
\begin{eqnarray}
\beta_1(t)=\beta_1\label{MP1}\\
\beta_2(t)=\beta_2\eta(t)\label{MP2}
\end{eqnarray}
and $\eta(t)$ is given in Eq.(\ref{eta}).
\begin{equation}
\ket{\psi(t)}=(\beta_1\ket{\mu}+\beta_2\eta(t)\ket{\nu})\otimes\ket{0}_e+\sum_k\ket{0}_1\ket{0}_2\otimes
c(t)_k\ket{1_k}_e\label{expr}
\end{equation}
The decoherence error come from the term
$\sum_k\ket{0}_1\ket{0}_2\otimes c(t)_k\ket{1_k}_e$. As shown in
Ref.\cite{Man}, by frequently measuring the environment, one can
remove the the term $|00\rangle$ and protect the initial state from
decoherence error, as long as one does not find a photon coming from the reservoir.
Intuitively speaking, such a frequent measurement works like a state
filter which removes $|00\rangle$ during the stage when its probability is
small. However, even though one can always remove the term with
$|00\rangle$ successfully by measurement, one can not protect the
initial state for a long time with the scheme because $\eta(t)$ decreases
significantly with time. Suppose the environment is measured after every time
interval $\Delta t$, and we continue to find no photon. At time $t$, the
state is changed into
\begin{equation}
|\psi(t)\rangle=(\beta_1\ket{\mu}+\beta_2
r(t)\ket{\nu})\otimes\ket{0}_e
\end{equation}
and
\begin{equation}
r(t)=\left[\eta(\Delta t)\right]^{t/\Delta t}.
\end{equation}
Since each measurement removes the photon in environment,
$\ket{\nu}\otimes\ket{0}_e$ restart the evolutionm from the initial
state again in each time interval. To protect two-qubit state of the
system more effectively, we can use the dynamical decoupling scheme
through applying $\pi$ pulses frequently instead of a filtration
scheme with frequent measurement. The main idea is, whenever the
state decays a little bit, i.e., $\eta (t)$ decreases a little bit
and the term with $|00\rangle$ appears with a small amplitude, a
$\pi$ pulse is applied and as a result, $\eta(t)$ will rise and the
amplitude of term with $|00\rangle$ decreases. That is to say, a
$\pi-$pulse does not simply remove the term with $|00\rangle$, it
changes the term with $|00\rangle$ back to the superradiant state.
Therefore, it differs from the state filtration of the measurement
based scheme - it   is really  a more effective scheme of state
recovery.

\section{Elimination of Decoherence with frequent $\pi$-phase pulses}

We show here that $\pi$-phase pulses can eliminate the decoherence on
the one hand and prevent amplitude changing of the superradiant
state on the other hand. A $\pi$-phase impulse take a phase-shift
operation as:
\begin{eqnarray}
\ket{0}\rightarrow -\ket{0} \;\ket{1}\rightarrow \ket{1}
\end{eqnarray}
Apply a $\pi-$pulse to each qubit, the two-qubit unitary operation
is then
\begin{eqnarray}\begin{array}{c}
\ket{1}_1\ket{0}_2\rightarrow\ket{1}_1\ket{0}_2\\
\ket{0}_1\ket{1}_2\rightarrow\ket{0}_1\ket{1}_2\\
\ket{0}_1\ket{0}_2\rightarrow -\ket{0}_1\ket{0}_2
\end{array}\end{eqnarray}
%
%
\subsection{Some iteration formulas}

Our method involves applying simultaneously $\pi$-pulses to each qubit.
To obtain the results of the method, we need some iteration formulas
first. Suppose the interval between two consequent impulses is
$\tau$ and $n=[{t \over \tau}]$, we can calculate the coefficients,
$r_1$ and $r_2$, by
\begin{widetext}
\begin{equation}
\begin{split}
\dot r_1(t)=&-\int _{n\tau}^t d\kappa f(t-\kappa)[\alpha^2_1
r_1(\kappa)+\alpha_1\alpha_2r_2(\kappa)]-\sum _{m=0}^{n-1}
(-1)^{n-m} \int _{m\tau}^{(m+1)\tau} d\kappa f(t-\kappa)[\alpha^2_1
r_1(\kappa)+\alpha_1\alpha_2r_2(\kappa)]  \\
\dot r_2(t)=&-\int _{n\tau}^t d\kappa f(t-\kappa)[\alpha^2_2
r_2(\kappa)+\alpha_1\alpha_2r_1(\kappa)]-\sum _{m=0}^{n-1}
(-1)^{n-m} \int _{m\tau}^{(m+1)\tau} d\kappa f(t-\kappa)[\alpha^2_2
r_2(\kappa)+\alpha_1\alpha_2r_1(\kappa)] \label{BB2}
\end{split}
\end{equation}
\end{widetext}
where we know $c (t)_k$, the coefficient of $\ket{1_k}_e$ in
Eqn.[\ref{expr}], changes the signal because of the two phase
impulses corresponding to $t=m\tau$(m is integer.).

Under the interaction with a Lorentzian spectral density, we get
the similar result as the free evolution, namely that there exist one dark
state $\ket{\mu}$ and one superradiant state $\ket{\nu}$. We next study
the condition that $\beta_1(t)$ and $\beta_2(t)$ should satisfy
between one time interval.

When $t$ is between $m\tau$ and $(m+1)\tau$.
\begin{eqnarray}
\beta_1(t)=\beta_1\label{MP3}\\
\ddot\beta_2(t)+\lambda\dot\beta_2(t)+R^2\beta_2(t)=0\label{MP4}
\end{eqnarray}
We know that the general solution of Eqn. \ref{MP4} is:
\begin{eqnarray}
\beta_2(t)=&e^{-{\lambda (t-m\tau) \over 2}}[A_m\cosh ({\Omega
(t-m\tau) \over 2}) \nonumber \\
&+B_m\sinh ({\Omega (t-m\tau) \over 2})] \label{general}
\end{eqnarray}
where $t\in [m\tau,(m+1)\tau]$. $A_m$ and $B_m$ are the constant
coefficients of the solution $\beta_2(t)$ at each time interval.
Our task now is to determine the relation of all $A_m$ and $B_m$.

At $t=(m+1)\tau$ using the boundary condition,
$\beta_2((m+1)\tau^-)=\beta_2((m+1)\tau^+)$ and
$\dot\beta_2((m+1)\tau^-)=-\dot\beta_2((m+1)\tau^+)$, we can get the
relation:
\begin{equation}
\begin{split}
A_{m+1}=&e^{-{\lambda\tau \over 2}}[A_m\cosh ({\Omega\tau \over
2})+B_m\sinh ({\Omega\tau \over 2})]\\
B_{m+1}=&-e^{-{\lambda\tau \over 2}}[A_m\sinh ({\Omega\tau \over
2})+B_m\cosh ({\Omega\tau \over 2})]\\
&+{2\lambda \over \Omega}e^{-{\lambda\tau \over 2}}[A_m\cosh
({\Omega\tau \over 2})+B_m\sinh ({\Omega\tau \over 2})]
\end{split}
\end{equation}
Here $\beta_2(m\tau^-)$ means the value of $\beta_2(m\tau)$ before
the pulse. And $\beta_2(m\tau^+)$ is the value of $\beta_2(m\tau)$
after the pulse.

\begin{figure}
  \includegraphics[width=160pt]{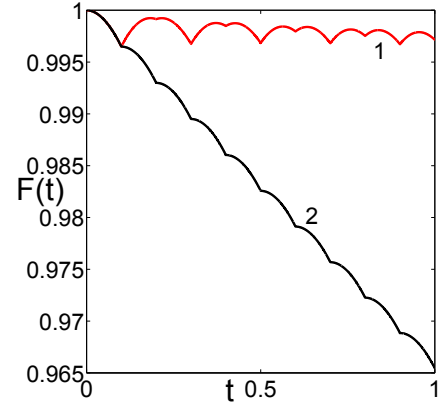}\\
  \caption{(Color online) Comparison of the fidelity evolution during $t=[0,\;1]$ of two methods for the superradiant state.
  We set $\tau=0.1$, $\lambda=2$ and $\Omega=1$.  Line 2 is for dynamical decoupling of this work and Line 1 is for QZE of Ref\cite{Man}.
  The fidelity only oscillates in a small range around a horizontal line in the dynamical decoupling method.}\label{3fig}
\end{figure}

We know that if the initial state is superradiant state,
$\beta_2(t)=1$, the initial value should be $A_0=1$ and
$B_0={\lambda \over \Omega}$. By using the relationship between
$A_m$, $B_m$, $A_{m+1}$ and $B_{m+1}$, we can get an analytical
function $\xi(t)$ which can express the fidelity
$\xi(t)=\sqrt{\bra{\nu}\rho_\nu(t)\ket{\nu}}$ of superradiant state.
\begin{eqnarray}
\xi(t)=&e^{-{\lambda (t-m\tau) \over 2}}[A_m\cosh ({\Omega (t-m\tau)
\over 2}) \nonumber\\
&+B_m\sinh ({\Omega (t-m\tau) \over 2})]
\end{eqnarray}
for $t\in[m\tau,(m+1)\tau]$. The fidelity $F(t)=\sqrt{\bra{\psi}\rho(t)\ket{\psi}}$ is given by
\begin{eqnarray}
\ket{\psi}=(\beta_1\ket{\mu}+\beta_2\ket{\nu})\\
F(t)=|\beta_1|^2+|\beta_2|^2\xi(t)
\end{eqnarray}
With this iteration formula for $\xi(t)$ above, we see tah
the fidelity oscillates about a horizontal line within a small range
after a pulse is applied, first increasing and then descending, as shown
in Fig [\ref{3fig}]. In contrast, the fidelity  based on the method of Quantum
Zeno Effect descends almost monotonously with time.

\subsection{Effect of finite duration of Double-$\pi$-Phase Operation}
In practice, one cannot set the duration pulse time to be infinitely
small. Here we consider the more realistic case that duration time
of double-$\pi$-phase is finite and we study show the effect on the
fidelity. We only consider the
sequential pulses method and we suppose the pulse duration  is $\tau \over
N$ in each time interval $\tau$.  We know that the decoherence
coefficient of Superradiant state is
Eqn. \ref{general} when $t\in [m\tau,(m+1-{1 \over N})\tau]$. So the effective Hamiltonian
for the sequential double-$\pi$-phase operator is $
H_{phase}={N\pi \over
\tau}(\sigma^+_1\sigma^-_1+\sigma^+_2\sigma^-_2)\sum_{m\geq 1}
[\Theta (m\tau-{1 \over N}\tau)-\Theta (m\tau)]$ and $\Theta(x)$ is
the step function. Under the Hamiltonian
$H_{all}=H_s+H_e+H_i+H_{phase}$, we can write down the state during the
time of $[(m-{1\over N})\tau , m\tau]$:
\begin{equation}
\begin{split}
\ket{\psi(t)}=&e^{-i{N\pi \over \tau}(t-(m-{1\over
N})\tau)}(r(t)_1\ket{1}_1\ket{0}_2\\
&+r(t)_2\ket{0}_1\ket{1}_2)\otimes\ket{0}_e\\
&+\sum_k\ket{0}_1\ket{0}_2\otimes c(t)_k\ket{1_k}_e
\end{split}
\end{equation}
In the same way as shown above, we also can get a integral equation describing
the coefficient $\dot r_1=-\int _0^t d\tau f'(t-\tau)[\alpha^2_1
r_1(\tau)+\alpha_1\alpha_2r_2(\tau)]$ and the new correlation
function of the duration time should be $f'(t)=W^2e^{(-\lambda
+i{N\pi \over \tau}) t}$. Using the boundary condition of
$t=(m-{1\over N})\tau$ and $t=m\tau$, we eliminate the coefficient
in the operator's duration time and get coupled relationships between $A_m$, $B_m$
and $A_{m+1}$, $B_{m+1}$(Eqn.\ref{general}).
\begin{equation}
\begin{split}
&e^{-{\lambda \over 2}(1-{1 \over N})\tau}[A_m cosh({\Omega \over
2}(1-{1 \over N})\tau)+B_m sinh({\Omega \over 2}(1-{1 \over
N})\tau)]\\
&=(1+\lambda{e^{\lambda\tau \over N}+1 \over 4i\gamma
-2\lambda})A_{m+1}-{e^{\lambda\tau \over N}+1 \over 4i\gamma
-2\lambda}\Omega B_{m+1} \\
&{\Omega\over \lambda}e^{-{\lambda \over 2}(1-{1 \over N})\tau}[A_m
sinh({\Omega \over 2}(1-{1 \over N})\tau)+B_m cosh({\Omega \over
2}(1-{1 \over N})\tau)\\
=&(1+e^{\lambda\tau\over N}+{\lambda \over
4i\gamma-2\lambda})A_{m+1}-({\Omega\over\lambda}e^{\lambda\tau\over
N}+{\Omega \over 4i\gamma-2\lambda})B_{m+1}
\end{split}\label{jj}
\end{equation}
where $\gamma={N\pi \over 2\tau}$.

The time evolution of the system is given by
$\ket{\psi(t)}=(\beta_1\ket{\mu}+\beta_2\xi(t)\ket{\nu})\otimes\ket{0}_e+\sum_k\ket{0}_1\ket{0}_2\otimes
c(t)_k\ket{1_k}_e$ and $\xi(t)$ is defined by Eqn. \ref{general}
with the new relation of $A_m$ and $B_m$ above. Also, the fidelity of
the state under decoherence with the original state is
\begin{eqnarray}
\ket{\psi}=(\beta_1\ket{\mu}+\beta_2\ket{\nu})\\
F(t)=|\beta_1|^2+|\beta_2|^2\xi(t)
\end{eqnarray}

\subsection{ Numerical Calculation of Dynamical Decoupling and
Free-evolution}

\begin{figure}
\begin{center}
  \includegraphics[width=160pt]{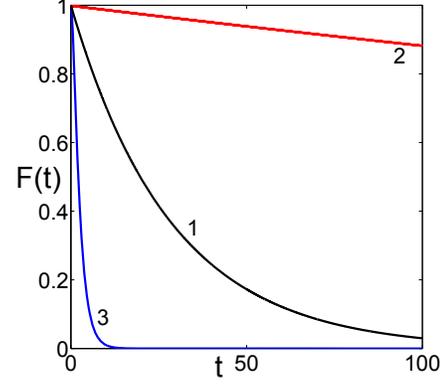}\\
  \caption{(Color online) The initial state is the superradiant state. Line 2 is the fidelity under sequential double-$\pi$-phase method
(Dynamical Decoupling). Line 3 is the fidelity under QZE. Line
1 is the fidelity of free-evolution under the interaction between
system and environment. From the result, we can easily find the
sequential-pulse method is much better than the QZE.}\label{1fig}
\end{center}
\end{figure}

\begin{figure}
\begin{center}
  \includegraphics[width=160pt]{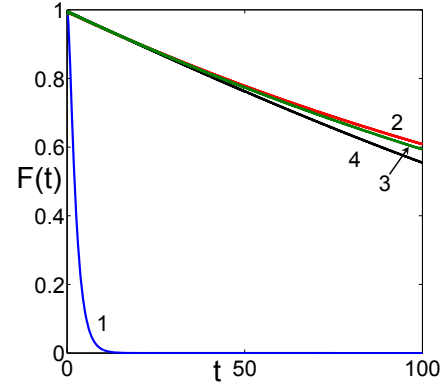}\\
  \caption{(Color online) The initial state is the superradiant state. Line 2 is the method with the instantaneous pulse. Line 4 is
for $N=20$ and line 3 is for $N=10$. Line 1 is for the
free-evolution.}\label{2fig}
\end{center}
\end{figure}

In case 1, we set $\tau=0.1$, $\lambda=2$ and $\Omega=1$ and we compare
the fidelity of the free-evolution, sequential double-$\pi$-phase
operator and random one. The result is shown in Fig[\ref{1fig}]. In
case 2, we set $\tau=0.2$, $\lambda=2$, $\Omega=1$, $N_1=10$ and
$N_2=20$ to compare how the duration time of pulses affect the
fidelity. The result is shown in Fig[\ref{2fig}].

\section{Concluding remark}
We present a strategy to protect the odd-parity states of two qubits
under 0 temperature environment by frequently applying the
$\pi-$pulses. Comparison between this method and the method based on
frequent measurement is done, it seems that the
frequent-$\pi-$pulses method is more effective in protecting the
states. Our result clearly shows that  the quantum Zeno effect
and the dynamical decoupling  can have rather {\em different} results 
when the operation frequency is finite, though the two methods give essentially the same results
 in the limit of infinite operation frequent as shown in Ref.~\cite{Facchi1,Facchi2,Busch}.
\\ {\bf Acknowledgments}
 XBW is supported by the National Natural
Science Foundation of China under Grant No.~60725416, the National
Fundamental Research Programs of China Grant No. 2007CB807900 and
2007CB807901, and China Hi-Tech Program  Grant No.~2006AA01Z420. LCK
acknowledges the financial support by the National Research
Foundation \& Ministry of Education.

\end{document}